# The curriculum prerequisite network: a tool for visualizing and analyzing academic curricula

Preston R. Aldrich
Department of Biological Sciences, Benedictine University, USA
Email: paldrich@ben.edu

**This article advances the prerequisite network as a means to visualize the hidden structure in an academic curriculum. Network technologies have been used for some time now in social analyses and more recently in biology in the areas of genomics and systems biology. Here I treat the curriculum as a complex system with nodes representing courses and links between nodes the course prerequisites as readily obtained from a course catalogue. The resulting curriculum prerequisite network can be rendered as a directed acyclic graph, which has certain desirable analytical features. The curriculum is seen as partitioned into numerous isolated course groupings, the size of the groups varying considerably. Individual courses are seen serving very different roles in the overall organization, such as information sources, hubs, and bridges. This network represents the intrinsic, hard-wired constraints on the flow of information in a curriculum, and is the organizational context within which learning occurs.**

## I. INTRODUCTION

All colleges and universities have a course catalogue. They are data-rich, and show course content and prerequisite mappings between courses. Unfortunately, course catalogues also are slow to yield information on relationships beyond one or two steps removed. Searchable electronic formats have improved their utility, yet the type of information we end up with is still very much the same as it probably has been since the course catalogue was invented. We manage to get the information faster, but we still read the catalogue as though it were a book.

The process of curriculum mapping [1–3] is a technical innovation in curricular studies that has gained popularity [4–7]. Curriculum mapping is a template-based procedure in which faculty identify content as it is delivered during a term, and then share this information with other faculty. Curriculum reform might then follow based on this global information. The approach is especially popular in K-12 education. Unfortunately, the information is typically organized in tables and spreadsheets that are not inherently easy to summarize or analyse.

The field of information visualization has grown substantially over the past two decades [8–11], fuelled in part by technical innovations in computing power and algorithms, and by the growing realization that an effective visual display of quantitative information can be critical to optimizing the usage of information. Networks are useful in this regard since they allow the visualization of broad-scale patterns and relationships without losing the fine-scale information on pairwise associations. Moreover, the mathematical field of graph theory provides numerous methods and metrics for the analysis of networks [12, 13], which are now a standard means of representing complex systems ranging from the internet to a living cell [14–17].

Networks play a prominent role in the visualization and analysis of knowledge systems. The concept map was introduced in 1990 [18], and is a network-based procedure for illuminating links between concepts. Directed acyclic graphs (or DAGs) are a type of directed network that does not contain cycles, i.e., one cannot leave a node and then later return to it, and these are critical to binary decision diagrams, certain Bayesian and neural networks, and other forms of machine learning [12, 19–21]. Neural networks are modelled after the brain, which is an extremely complex set of neurons interlinked as a network [22]. Collective knowledge networks such as Wikipedia [23] reside on the network of web pages known as the World Wide Web, itself hosted on the network infrastructure of the internet [13]. And collective academic knowledge has been studied extensively through the fields of scientometrics and bibliometrics, where data mining procedures are used to extract information from the scientific literature yielding broad-scale network visualizations of scientific knowledge and the process of science [24–26].

## II. RELATED WORK

Curriculum visualization and analysis has been attempted using several different methods, and in most cases the emphasis has been on understanding curricular structure in order to modify and improve it. As early as 1984, student enrolment data in Drexel

University's School of Library and Information Science were analyzed using multidimensional scaling to generate a two-dimensional map of the top forty two elective courses [27]. By 1990, concept maps were being used to improve a science curriculum [28]. By 2007, there began an upsurge in interest regarding curriculum visualization methods with most reports appearing in conference proceedings, and all having an applied emphasis such as seeking to enhance student experience during course selection and facilitating discussions amongst faculty regarding curricular revisions [29–34]. For example, the computer science program at Ball State University was visualized as a digraph that was used during advising and recruitment events to explain the curriculum [30]. And there has been a social network study of Chinese academic web pages [35]. To my knowledge, there is no published study to date in which a CPN built for an entire college or university curriculum has been the object of study. The subject is interesting because it regards the fundamental constraints on information flow within an academic institution.

III. RESEARCH QUESTIONS

Here, I explore the Curriculum Prerequisite Network, or CPN, which is a network view of the knowledge system contained in a university course catalogue. Nodes (or vertices in graph theory) represent courses, and directed links (or arcs) between courses represent prerequisite requirements. The entire system is a directed graph or digraph.

Research questions fall into three categories. (1) The first question is methodological; what issues are encountered when coding a CPN from a university course catalogue? Registrars are notoriously meticulous about maintaining consistent catalogue syntax, so it is a reasonable expectation that some of the re-coding might be automated. And, what other catalogue elements beyond the prerequisite link require attention? (2) The second question concerns the global topology of a CPN. In its native state, is it a DAG? Of how many different parts is it comprised, how do they fit together, and how might this influence the flow of information? (3) The third question relates to the role played by individual courses in the curriculum. From a topological perspective, how are courses differentiated with respect to their position within a curriculum, how might we measure this in a graph theoretic context, and how might their roles influence the flow of information? The overall goal here is not to provide an exhaustive appraisal of these questions, but to begin developing the perspective of viewing CPNs as a proper object of study in their own right.

IV. METHODS

*A. From catalogue to network*

The course catalogue of Benedictine University, Lisle, IL, USA, was used for this study. Benedictine University is an independent, Catholic, comprehensive, 501(c)(3) institution of higher education, category 17 (DRU: Doctoral/Research Universities) of the Carnegie BASIC2010 classification. The main campus is in the Chicago metropolitan area, and serves a racially, ethnically, and religiously diverse student population of 7,434 students (4,455 undergraduates and 3,543 graduate). It offers fifty three undergraduate majors, twelve masters programs, two Ph.D. programs, and an Ed.D. program. A primary strength of the institution is its well-established and highly regarded science programs.

Information from the undergraduate course catalogue for 2009-2010 was obtained from the university website. Catalogue information was scraped using Python and the Beautiful Soup library which creates a parse tree for HTML documents. Data cleaning involved additional rounds of Python scripting in which extraneous information was removed and markup added to aid precise extraction. The information ultimately was converted into network format using NetworkX (version 1.8), a Python package for the analysis of complex networks provided by the Los Alamos National Laboratory [36].

As an example, Figure 1 shows a portion of a hypothetical course catalogue along with the CPN that would result. The dashed arcs are not coded from the catalogue shown, but are included in the network for heuristic purposes to illustrate mappings that would introduce cycles into the CPN. A cycle in a graph allows one to travel away from a node and then at some point return back to it. Although students sometimes find themselves repeating a course, it should be because of poor performance in a prior semester, not because of curricular bindings. The three-membered cycle shown (involving the hypothetical BIOL 100) is unlikely to arise except by error, since it excludes enrolment in the member courses. A student could enrol in BIOL 100 only if s/he had already completed BIOL 110, CHEM 100, – and BIOL 100. Cycles arising from corequisites are addressed in the general rules and conventions that follow:

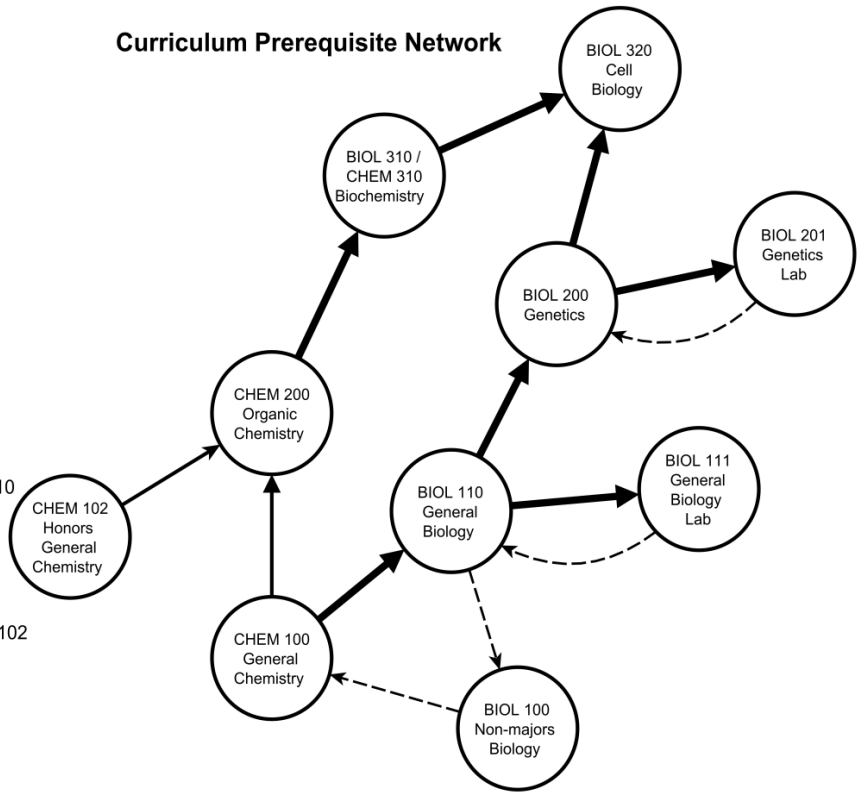

Fig. 1: Hypothetical course catalogue and the associated CPN. Rendered using yEd (http://www.yworks.com).

*Prerequisites*. The prerequisite binding – course A must be taken before course B – establishes a parent-daughter, predecessor-successor relationship between the courses, one that is readily modelled as a directed graph. Although one could specify B → A, where B is referencing information obtained previously in course A, I chose the alternate method that better represents the flow of information from A into B, the arc A → B. If more than one prerequisite was specified as mandatory, then equal and full weights were applied to each arc. For example, if both course A and B were required before one could take course C, then arcs A → C and B → C were both given a weight of 1.0. However, if prerequisite rules presented an option such that, for example, either course A or B (but not both) was required before C, then the total weight of 1.0 was distributed equally across arcs (e.g., A → C, 0.5; B → C, 0.5).

*Corequisites*. A hard corequisite binding – coregistration in course X – establishes a symmetric relationship between courses, particularly if both courses reference the other as a corequisite. Here, it generally is assumed that a student will enroll for both courses A and B in the same term, as is typical of lecture/lab combinations in the sciences, or more broadly when a theory course is temporally bound to a practical applications course. We can represent this symmetric binding with a bidirectional arc between the courses in the CPN (A ←→ B, Figure 1), with equal weights in both directions (A → B, 1.0; B → A, 1.0). However, this bidirectional edge introduces a 2-member cycle, and the CPN no longer can be treated as a directed, acyclic graph (DAG).

A soft corequisite binding – credit or coregistration in course X – establishes a slightly asymmetric relationship between courses, particularly if only one course of the two names the other as a corequisite. Although most students likely will take A and B together, the option exists to take course A first and B later. This allowance of temporal priority also suggests a certain amount of conceptual priority. One might ask, if a student were to take the courses in different terms, which should come first? I expect most would agree that a lecture should precede a lab, not the other way around. By this argument, one might treat some corequisite courses as conceptually, if not temporally, prior in the curriculum.

In the Benedictine University catalogue, this was a reasonable interpretation since all corequisite pairings contained one course (A) that served as a 'lecture' or 'theory' course and the other (B) as a 'laboratory' or 'applied' course. Consequently, I elected to encode all corequisite relationships as prerequisite bindings (B → A), thus preserving the DAG structure; this was not critical to the analyses performed in the present study. The hypothetical example in Figure 1 shows two instances of corequisite bindings between a lecture and a lab, wherein a cycle is introduced if a bidirectional arc is used, whereas DAG structure is preserved if only the lab carries the co(pre)requisite.

*Cross-listings.* The cross-listing establishes an equivalency between courses A and B. One might choose to keep courses A and B separate in a CPN, if they were indeed distinct courses that served a similar but not identical role in a curriculum. A more literal reading of the catalogue would be to treat them as one and the same course. I chose the latter option, and merged cross-listed course nodes creating a composite node (A/B) inheriting all of the bindings of the original nodes.

*Hard-wired vs. soft-wired relationships.* The course catalogue also contained several diffuse (soft-wired) bindings between courses, such as prerequisite rules that specify Junior or Senior standing before enrolment in a course. Although such bindings could be modelled in a CPN, I chose for this exercise to focus exclusively on bindings that were explicitly stated for specific courses (hard-wired).

*B. Analysis of network topology*

The CPN was initially constructed as a directed graph. Python/NetworkX [36] was used to assess whether it also was a directed, acyclic graph (DAG). Cycles were removed when detected, resulting in a CPN with a DAG architecture. The DAG-CPN (hereafter, CPN) was then analyzed using standard graph metrics with NetworkX [36] and Gephi [37].

The degree of a node ($k$) in a directed graph is the sum of the in-coming arcs ($k_{in}$, or in-degree) and the out-going arcs ($k_{out}$, or out-degree). Weighted degree [38] was evaluated as:

$$k_i = \sum_{j=1}^{N} a_{ij} w_{ij} \quad \quad \text{1]}$$

where $k_i$ is the weighted degree of node $i$, and $a_{ij} = 1$ if there exists a connection between nodes $i$ and $j$, and $a_{ij} = 0$ otherwise. The strength of the connection is denoted by the weight $w_{ij}$, which is 1.0 for all links when evaluating an unweighted node degree. A node with a very high degree is termed a 'hub' [13], and may be important in the channelling of information.

Betweenness centrality ($b_i$) is another popular index, measuring the extent to which a node lies on the shortest paths between the other nodes in a graph. As such, it speaks to the broader-scale traversability of a network, and to the role played by individual nodes in that level of connectivity. A node with high betweenness tends to act as a bridge or conduit between large but otherwise isolated regions of a network. The unweighted implementation in Gephi was used [39]. By definition, the distance between two nodes is infinite if they reside in separate connected components (see below), so betweenness coefficients were evaluated only for nodes in the largest connected component.

Analysis of connected components allowed appraisal of the extent to which the CPN was subdivided into different, disconnected groupings of courses. Weakly connected components were detected using Gephi and a depth-first search algorithm [40]. In a directed graph, a weakly connected component is a maximal connected subgraph, or largest set of nodes interlinked without regard to the directionality of the arcs. A strongly connected component is defined as a set of nodes that are each reachable from any node in the set while paying attention to arc directionality [12]. In a DAG, there are no two nodes that are both reachable starting from the other node, so the number of strongly connected components equals the number of nodes in the graph, which does not interest us here.

V. RESULTS

*A. DAG topology*

Analysis of the CPN using NetworkX indicated that the catalogue, interpreted literally, contained coding for 109 cycles, and so was not a DAG. All of these cycles were due to corequisite bindings between a lecture or theory course (A) and a laboratory or applications course (B). In order to rectify this, the lab prerequisite was removed from each lecture course (though most did not have one), while the labs retained the lecture prerequisite (B → A). Analysis confirmed that the modified CPN had a DAG architecture. All subsequent analyses were performed on this DAG-CPN (hereafter, CPN).

*B. Global curricular topology*

Summary metrics are reported in Table 1. The full undergraduate CPN of Benedictine University is shown in Figure 2. There were 1,097 nodes (courses) in the 2009-10 catalogue, with 770 arcs (prerequisite bindings). A total of 92 of the nodes (8.4%) were composite courses resulting from cross-listing. The average shortest path connecting nodes in the largest connected component involved 2.48 steps, which fully binds at least three courses in a sequence. The longest of the shortest paths (diameter, 6) bound seven courses.

Table 1. Centrality and traversability metrics for the full CPN and its largest connected component.

| Metrics | Full CPN | Largest connected component |
| --- | --- | --- |
| nodes | 1,097 | 328 |
| arcs | 770 | 530 |
| density | 0.00128 | 0.00988 |
| weakly connected components | 599 | 1 |
| degree | 1.40 | 3.23 |
|    in-degree | 0.70 | 1.62 |
|    out-degree | 0.70 | 1.62 |
| weighted degree | 1.10 | 2.44 |
|    weighted in-degree | 0.55 | 1.22 |
|    weighted out-degree | 0.55 | 1.22 |
| diameter | -- | 6 |
| characteristic path length | -- | 2.48 |
| betweenness centrality | -- | 0.000103 |

[a] Standard deviation in parentheses.

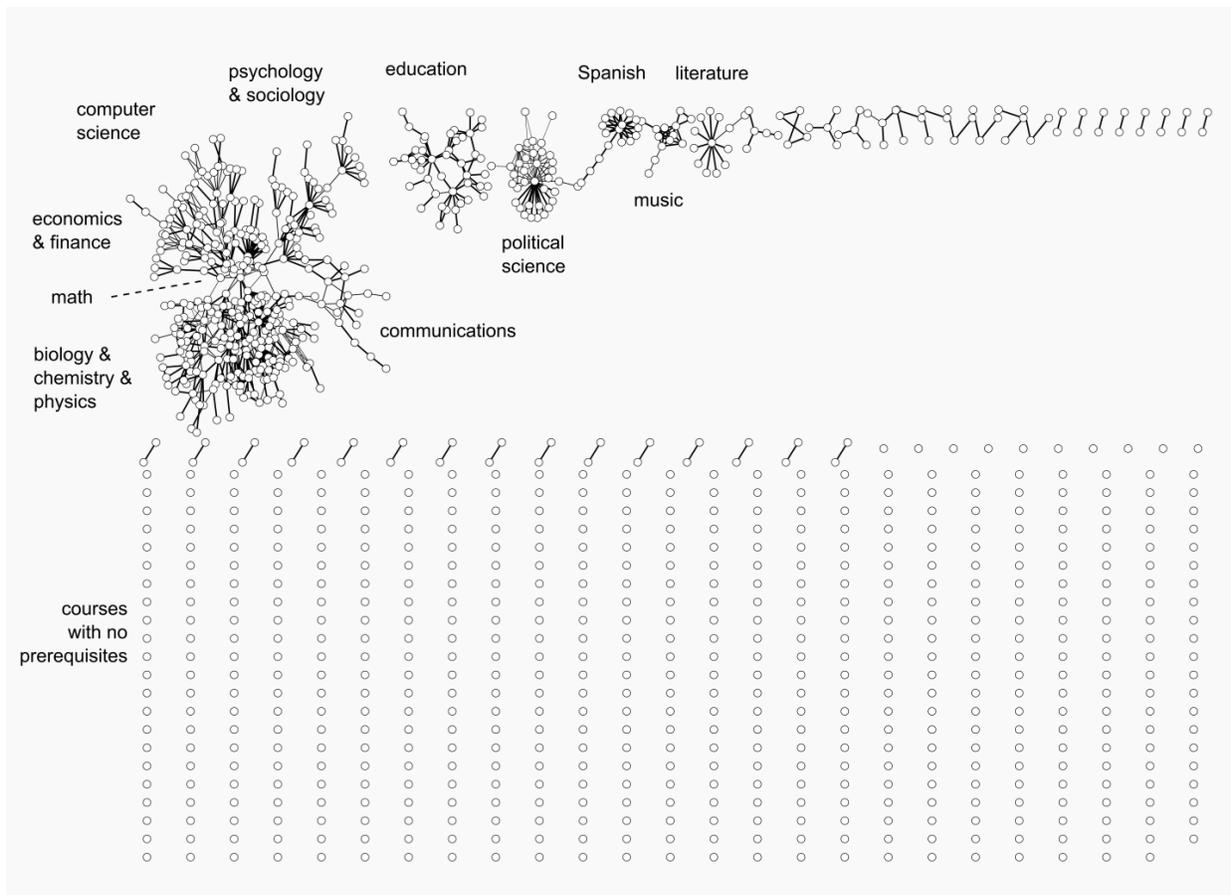

Fig. 2: Entire 2009-10 undergraduate curriculum of Benedictine University as a Curriculum Prerequisite Network (CPN; rendered using Pajek [41])

The connectivity of the CPN was highly heterogeneous. The average node connected to the weighted equivalent of one other node (weighted $k$ = 1.10), with half of these connections incoming and half outgoing. However, this 'average' node was poorly represented in the actual CPN (Figure 2). Over half of the nodes ($n = 559$, 51.0%) had a degree of zero, and nearly a third of the nodes ($n = 328$, 29.9%) resided in the largest connected component, where the mean degree was (weighted $k = 2.44$) over twice that in the CPN overall (Table 1, Figure 1). The largest connected component contained over half of the total arcs ($n = 530$, 68.8%). The full CPN contained 599 weakly connected components.

*C. Local topology and individual courses*

The courses varied widely in the metrics examined (Tables 2, 3, Figure 3). Courses outside the natural sciences were well-represented (60%) in the top ten list for out-degree centrality, but not in the betweenness centrality top ten list. There was a significant correlation between weighted node degree and betweenness (Spearman rank order correlation $rho = 0.66$, $P < 0.001$, Figure 4). Out-degree centrality is a local graph metric measuring, in a CPN, the number of courses that call a particular course as a prerequisite. By contrast, betweenness centrality is a global metric that is sensitive to the overall tranversability of a graph, and to the role played by an individual course in facilitating that traffic.

Table 2. Top 10 Benedictine University undergraduate courses based on weighted out-degree centrality (entire graph considered).

| Course number | Course name | Degree centrality |
|---|---|---|
| BIOL 108[*] | Principles of Biology | 29.0 |
| PLSC 102 | American Government | 19.2 |
| BIOL 109[*] | Principles of Biology Lab | 14.0 |
| BIOL 258 | Human Physiology | 13.0 |
| SPAN 211 | Intermediate Grammar & Composition | 12.0 |
| CHEM 123 | General Chemistry II | 12.0 |
| PSYC 100 | Survey of Psychology | 11.5 |
| SPAN 212 | Intermediate Oral Communication | 11.0 |
| EDUC 205 | History and Philosophy of Education | 11.0 |
| LITR 100 | Introduction to Literary Analysis | 10.5 |

[*] This catalogue was in transition for BIOL 108, which was splitting into BIOL 197 and 198, and BIOL 109 which was converting to BIOL 199. All old names were retained for this analysis.

Table 3. Top 10 Benedictine University undergraduate courses based on unweighted betweenness centrality (only largest connected component considered).

| Course number | Course name | Betweenness centrality |
|---|---|---|
| CIS 200 / CMSC 200 | Computer Programming | 0.0021 |
| CHEM 123 | General Chemistry II | 0.0014 |
| CHEM 113 | General Chemistry I | 0.0014 |
| MATH 110 | College Algebra | 0.0013 |
| CIS 274 / CMSC 274 | OO Design and Programming | 0.0012 |
| BIOL 340 | Cell Biology | 0.0011 |
| MATH 210 | Calculus with Analytics I | 0.0011 |
| MATH 111 | College Trigonometry | 0.0011 |
| CHEM 243 | Organic Chemistry I Lab | 0.0011 |
| NTSC 152 | Natural Science Interdisciplinary Lab II | 0.0011 |

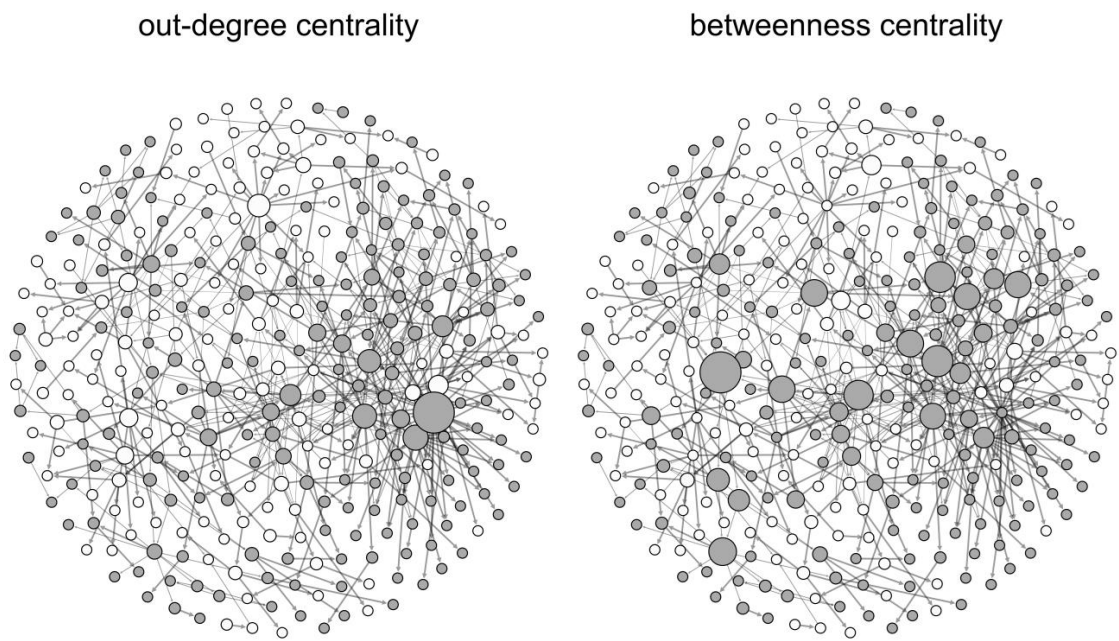

Fig. 3: Largest connected component of the Benedictine University undergraduate CPN (dark nodes, College of Science courses; node size represents out-degree centrality (left) and betweenness centrality (right; using Gephi).

The general biology course, BIOL 108, occupied an influential position in the curriculum as a hub since so many other courses called it as a direct prerequisite ($k = 29.0$). Yet, it was not centrally located in the curriculum since it was the first course taken in a variety of different learning pathways involving the life sciences. BIOL 108, in this way, acted as a primary information source. But as a consequence of its early position in the curriculum, it had a fairly unimpressive betweenness centrality (Figure 4, 5). PSYC 100 Survey of Psychology had a similar profile.

Several of the highly connected (high-degree) courses outside the natural sciences, e.g., PLSC 102 American Government, resided in the smaller connected components where they served as a prerequisite to many courses within their own department. However, they were not on very many of the shortest paths between courses in the overall curriculum due to their isolation, and so had low betweenness coefficients.

Bridge courses have high betweenness centralities (Table 3), as they channel information between different segments of the curriculum. The composite course CIS 200/CMSC 200 Computer Programming had the highest value. CHEM 123 General Chemistry II had the next highest value, and was the only course appearing in both top ten lists (Tables 2, 3). All of the courses with highest betweenness were College of Science courses, 60% either mathematics or chemistry.

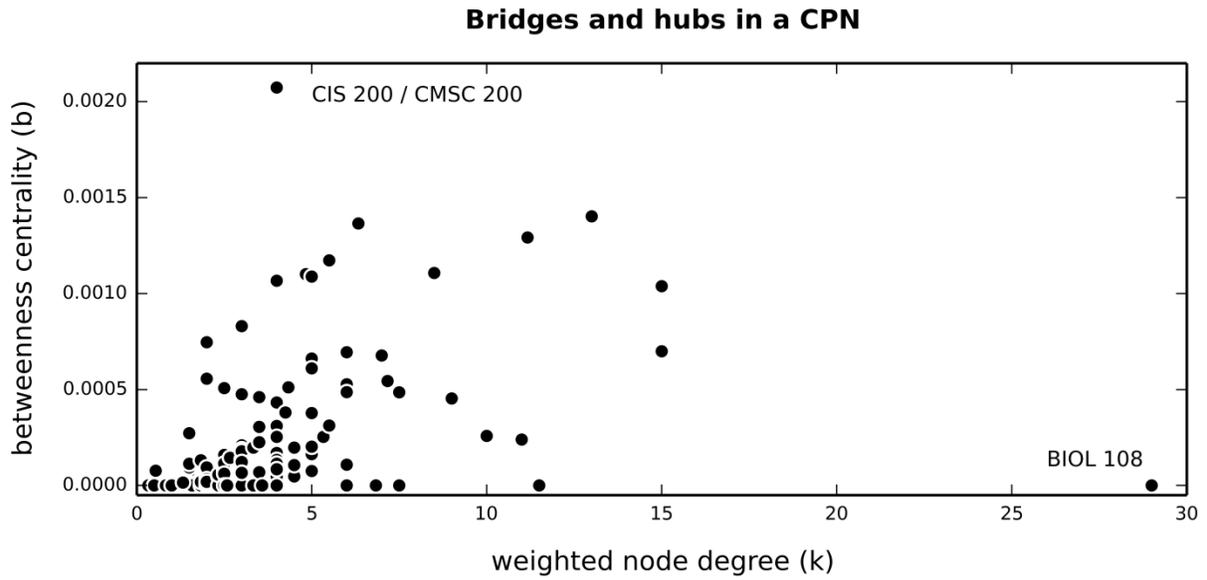

Fig. 4: Weighted node degree versus betweenness centrality.

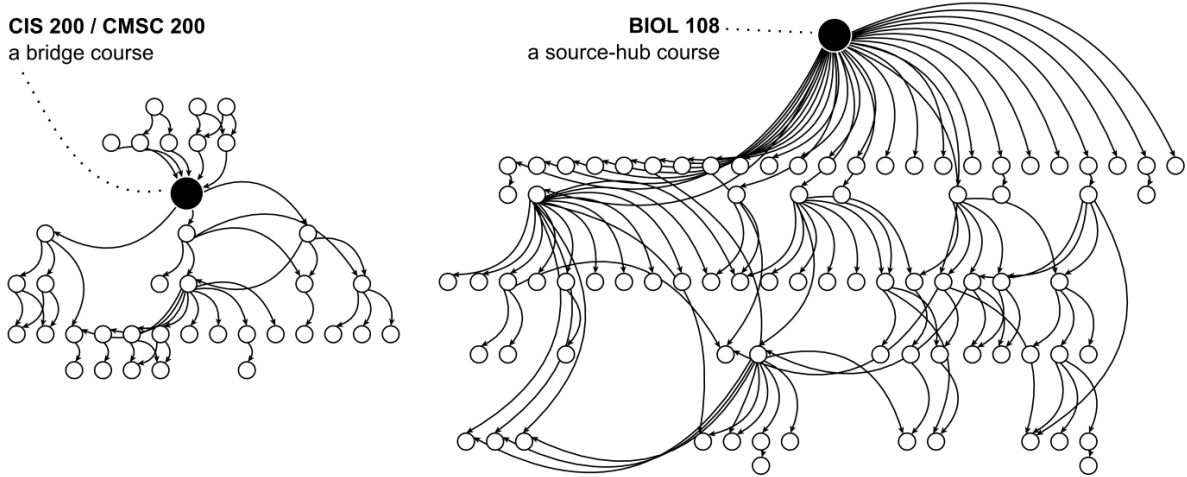

Fig. 5: A bridge and a source-hub (rendered using yEd).

## VI. Discussion and conclusions

### A. CPN construction

It was a reasonable task taking the information in a catalogue and translating it into a network. The syntax of the catalogue was highly regular, and most of the information was readily coded as pairwise bindings between courses. There exist some aspects of the catalogue that were not captured by this CPN such as the soft-wired rules, but these could be addressed if it was deemed important. It could be argued there was some sacrifice of reality in re-coding the corequisite relations as prerequisites, this to save the analyticity of the DAG. But if the judgment was made that losing the DAG structure was warranted in order to retain corequisite relations with the cycles that they imply, there are still many interesting things that could be done with a CPN as a simple directed graph.

### B. CPN topology

The Benedictine University undergraduate CPN for 2009-10 displayed several interesting topological features that were not immediately evident from a casual reading of the course catalogue.

Roughly half of the curriculum was bound together in a largest connected component in which the courses fed into one another through a complex set of prerequisite relationships. This largest component consisted predominantly of courses offered by the College of Science (190 COS, 57.9%; 138 other, 42.1%). Benedictine University has a tradition of excellence in the sciences, and so it is not surprising that science courses would occupy a central position in the curriculum; natural science students represent 36% of all declared undergraduate majors. Not surprisingly, the longest of all the shortest paths within this largest component consisted of only six steps (involving seven nodes), which translates to seven sequential semesters or 3.5 academic years. This makes sense given the standard graduation target is four years for a baccalaureate degree.

There were a few connected components of much smaller size though still comprised of more than one node. Each group represented the offerings from a single department. Several of the components were radially symmetric with several courses calling a single, key course as a prerequisite, and these had some of the highest node degrees in the study.

Over half of the nodes in the total curriculum ($n = 559$, 51.0%) had a degree of zero; these were courses that carried no prerequisite requirement. Many, though not all, of these courses represented elective credit that students might take outside their major, and such courses cannot be burdened with prerequisites if they are to serve this role in the curriculum. Though it does bring into focus one fact; a substantial part of a curriculum might not cohere. A lack of prerequisite/corequisite bindings for a large number of courses indicates that there are several degrees of freedom in the system. Given the importance of academic freedom in the delivery of classroom content, this is not automatically a negative quality but perhaps even an asset. And, based on my own experiences at the institution, the faculty do work to cohere their own courses with others offered on campus, whether or not their courses are bound by prerequisites. But it does suggest that there might be several opportunities to introduce coherence in this collection of unbound courses without introducing prerequisite requirements, should that be seen as desirable.

### C. Roles of courses in the curriculum

These results show how courses can differ in their roles in a curriculum. While most courses were not distinguished by topology, a few stood out and played a significant role in controlling the flow of information.

Hub prerequisite courses are highly-connected courses that are called by numerous down-stream courses. The case shown here was of the general biology course that also served the role of key source due to its position early in the curriculum.

Some courses acted as bridges between regions of the curriculum that were otherwise not integrated. The computer programming course served this role, bridging several early courses with several later ones. Apparently computer programming is not an entry-level course, but it is sufficiently critical at a point so that many computer courses eventually call it as a prerequisite. As such, the course is an information bridge. That it is cross-listed (CIS/CMCS 200) contributes to its effectiveness at binding different segments of the curriculum.

### D. Prospects

CPNs are useful tools for the viewing and analysis of curricula. The course catalogue is a rich resource of infrastructural information that, when observed from the right vantage point, can illuminate interesting higher-order organizational attributes of an academic institution. Such data mining approaches as undertaken here are increasingly common in higher education [42], and given the recent interest in alternative curriculum visualization methods [27–35], it is unlikely we will continue to interact with our course catalogues the same way for much longer. The present work places into focus the fact that the CPN is an interesting and worthy object of academic study, as it represents the collective constraints in a curriculum on the flow of information and forms the framework within which academics deliver course content.